# An asymmetric elastic metamaterial model for elastic wave cloaking

H.K. Zhang, Y. Chen, X.N. Liu[*] and G.K. Hu[*]

Key Laboratory of Dynamics and Control of Flight Vehicle, Ministry of Education, School of Aerospace Engineering, Beijing Institute of Technology, Beijing 100081, China

**Abstract:**

Elastic material with its elastic tensor losing minor symmetry is considered impossible without introducing artificially body torque. Here we demonstrate the feasibility of such material by introducing rotational resonance, the amplified rotational inertia of the microstructure during dynamical loading breaks naturally the shear stress symmetry, without resorting to external body torque or any other active means. This concept is illustrated through a realistic mass-spring model together with analytical homogenization technique and band structure analysis. It is also proven that this metamaterial model can be deliberately tuned to meet the material requirement defined by transformation method for full control of elastic wave, and the relation bridging the microstructure and the desired wave functionality is explicitly given. Application of this asymmetric metamaterial to design elastic wave cloak is demonstrated and validated by numerical simulation. The study paves the way for material design used to construct the transformation media for controlling elastic wave and related devices.

**Key Words**: Asymmetric metamaterials, minor symmetry, elastic cloak, mass-spring lattice, elastic wave

---

[*] Corresponding authors: liuxn@bit.edu.cn, hugeng@bit.edu.cn

# 1. Introduction

The elastic tensor of classical materials possesses both major and minor symmetry, the former is required by the assumed reciprocity, and the latter (hence the symmetry of shear stress) comes from the fact that the angular momentum conservation can be ignored in the absence of body torque (Fung, 1965). This theory is recognized a tremendous success for engineering structure design. In the recent decade, the emergence of elastic metamaterials extends the border of the elastic materials by allowing exotic material constants, such as negative mass density (Liu et al., 2000; Yao et al., 2010), modulus (Fang et al., 2006; Liu et al., 2011) or both negative (Zhu et al., 2014), the elastic tensor still respects the major and minor symmetry. On the other hand, breaking minor symmetry is well known for the higher order continuum, e.g. micropolar or Cosserat media, for which the material particle cannot be considered as infinitesimal and admits independent rotation and body torque (Eringen, 1999; Nowacki, 1986). This makes the stress and strain tensors asymmetric, meanwhile additional equations concerning the angular momentum conservation and higher order stress and strain measures must be accompanied to render the whole problem rigorous. The standalone Cauchy media without minor symmetry (henceforth called asymmetric metamaterial (AMM)) is considered however impossible if there exists no external body torque, since the objectivity is obviously violated.

Although not available in practice yet, these fictitious AMMs are among very few models (together with Willis material model) which are able to be impedance-matched with Cauchy media, an extremely important property for elastic wave mitigation. They are used for a long time in numerical simulation to construct artificial perfect matching layer (PML) for elastic wave to mimic an infinite space without reflection (Zheng and Huang, 2002; Chang et al., 2014). In the past two decades, stimulated by the rapid development of metamaterial techniques (Christensen et al., 2015; Ma and Sheng, 2016), the concept is systemized into the so-called transformation method (Greenleaf et al., 2003; Leonhardt, 2006; Pendry et al., 2006) with which various physical fields can be imagined to be freely manipulated, e.g. electromagnetic wave (Schurig et al., 2006), acoustic wave (Cummer and Schurig, 2007), thermal conduction (Han et al., 2013) and elastic wave (Milton et al., 2006) etc., provided that the transformation induced material property (usually not find in nature) can be designed. The most challenging application of the transformation method is the design of cloak that makes a region undetectable from the surrounding field. For elastic wave, it is demonstrated that a perfect elastic cloak resorts to the AMMs (Brun et al., 2009) or the Willis media (Milton et al., 2006), depending on the choice of gauge monitoring the migration of displacement vector from the virtual space to the transformed space (Norris and Shuvalov, 2011). Though the asymmetric elastic tensor is directly available for micropolar materials, recent studies show that the full Cosserat theory is not form invariant just like Cauchy model under a general space mapping, this makes the design of elastic wave cloak in obscurity. There is a reduced Cosserat theory originally introduced by Schwartz (Schwartz et al., 1984) for understanding waves in aggregate of packed solid spheres. Contrary to the full version, the material reacts only to the particle rotation relative to the host, but there is no direct interaction trying to reduce the relative rotation of two particles. Thus the material may have asymmetric stress but zero couple stress. Grekova et al. (2009) argued that this reduced Cosserat medium could be further recast into an AMM by eliminating the particle rotation. However, the discussion is neither associated to the transformation theory nor the way to realize this AMM. Theoretically, an elastic cloak can also be realized within framework of the small-on-large theory with a pre-stressed hyperelastic material (Norris and Parnell, 2012), the tangent moduli of a semi-linear hyperelastic material may satisfy the material requirement defined by the transformation theory, which has no minor symmetry. But its feasibility is restricted to the semi-linear materials not available at present. Another route to design elastic cloak can be followed with Willis' materials, however their microstructure design is far from mature. More recently Nassar et al. (Nassar et al., 2018; Nassar et al., 2019) proposed an active mass-spring network to mimic the AMMs by using grounding mechanism, originally from the torque spring proposed by Milton (Milton and Seppecher, 2008; Vasquez et al., 2011). Their work makes an important step towards elastic cloak design. However, a challenge still remains: are the passive AMMs feasible without introducing external body torque or couple stress?

With such question in mind, let's imagine the following scenario: if the microstructure is cleverly designed inside a material element, the amplified rotational inertia by rotational resonance of the microstructure during loading remains local and induces no couple stress, this rotational effect should be balanced by asymmetric stresses. In this way, AMMs could be conceived in principle. Following this vein, we will detail the idea in this study, and demonstrate the feasibility of such metamaterial with the design of an elastic cloak. The paper goes as follows: the mechanism and wave property of AMMs are explained in Section 2; the detailed microstructure design and validation along with the homogenization and the inverse microstructure parametrization for wave functionalities are given in Section 3; the design and numerical validation of an elastic cloak are provided in Section 4.

**2. Elastic metamaterial without minor symmetry**

When the minor symmetry of an elastic material is discarded, the displacement gradient is taken directly as strain measure, and the linear elastic constitutive equation reads

$$\sigma_{ij} = C_{ijkl}\varepsilon_{kl} = C_{ijkl}u_{l,k}, \tag{1}$$

where the stress $\sigma_{ij}$ and strain $\varepsilon_{ij}$ are no longer symmetric tensor, and $C_{ijkl} \neq C_{ijlk} \neq C_{jilk}$. However, the major symmetry, $C_{ijkl} = C_{klij}$, should hold since the material is assumed to be reciprocal. To draw a physical picture of how this material behaves, let us consider two dimensional (2D) isotropic case, for which the asymmetric elastic tensor can be written as (Liu and Hu, 2005)

$$C_{ijkl} = \lambda \delta_{ij}\delta_{kl} + (\mu+\kappa)\delta_{ik}\delta_{jl} + (\mu-\kappa)\delta_{jk}\delta_{il}, \tag{2}$$

where $\delta_{ij}$ is Kronecker delta, $\lambda$ and $\mu$ are the traditional Lame's constants and $\kappa$ is the skew-symmetric shear modulus. Splitting an asymmetric tensor as $s_{ij} = s_{(ij)} + s_{<ij>}$, where $s_{(ij)} = (s_{ij} + s_{ji})/2$ and $s_{<ij>} = (s_{ij} - s_{ji})/2$ are the symmetric and skew-symmetric parts, respectively, Eqs. (1) and (2) give the relations between the shear part of stress and strain ($i \neq j$) by

$$\sigma_{(ij)} = 2\mu\varepsilon_{(ij)}, \qquad \sigma_{<ij>} = 2\kappa\varepsilon_{<ij>}, \tag{3}$$

and the implied physical mechanism is sketched by Fig. 1(a). For an elastic media without minor symmetry ($\kappa \neq 0$), a pure rigid rotation $u_i = R_{ij}x_j$ ($R_{ij}$ is the rotation matrix, skew-symmetric for small deformation) will induce internal energy, and the restoration force on the element is just $\sigma_{<ij>}$. Note that this is different from the Cosserat model in which the rotation is an independent degree of freedom (DOF). For a pure continuum without microstructure or external source torque, this constitutive behavior is not allowed under the requirement of objectivity. However, as shown in Fig. 1(b), if we consider a metamaterial unit cell with a hidden rotational microstructure, when rotation resonance is excited and out-of-phase relative to the observable part, a skew-symmetric stress must be detected on the cell boundary to balance the overall angular momentum (Fig. 1(c)), hence the metamaterial will effectively lose the minor symmetry.

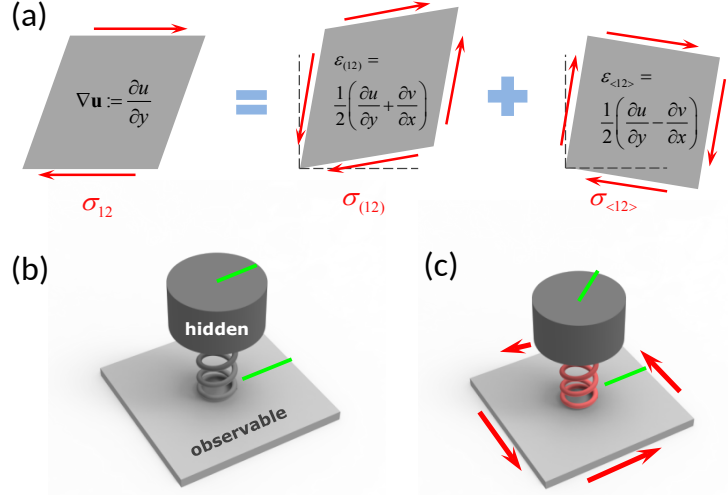

Fig.1. (a) The asymmetric shear stress and deformation is superposition of the symmetric part and skew-symmetric part; (b) no skew-symmetric stress if there is no relative rotation between the hidden and observable parts, as indicated by the green lines; (c) near the resonance, the torque in the spring demands skew-symmetric stress on the observable element, renders the stress tensor asymmetric.

Let us consider a concrete example of asymmetric elasticity arising from the coordinate transformation of wave equations of traditional Cauchy media. Under a curvilinear coordinate mapping from the virtual space $\mathbf{x}'$ to the physical space $\mathbf{x}$, and choosing the gauge that displacements parallelly migrate between original domain $\Omega'$ and deformed domain $\Omega$, i.e. $\mathbf{u}' = \mathbf{u}$, the transformation of wave equations for a traditional elastic media $(\rho_0, \mathbf{C}^{sym})$ is as the following (Norris, 2011)

$$\begin{cases} \nabla' \cdot \boldsymbol{\sigma}' = \omega^2 \rho_0 \mathbf{u}' \\ \boldsymbol{\sigma}' = \mathbf{C}^{sym} : \nabla' \mathbf{u}' \end{cases} \rightarrow \begin{cases} \nabla \cdot \boldsymbol{\sigma} = \omega^2 \rho \mathbf{u} \\ \boldsymbol{\sigma} = \mathbf{C} : \nabla \mathbf{u} \end{cases} \tag{4}$$

where the transformed stress and material parameters are provided respectively by

$$\sigma_{ij} = J^{-1} F_{ip} \sigma'_{pj}, \quad \rho = J^{-1} \rho_0, \quad C_{ijkl} = J^{-1} F_{ip} C^{sym}_{pjql} F_{kq}, \tag{5}$$

with $F_{ij} = \partial x_i / \partial x'_j$ and $J = \det \mathbf{F}$ being separately the deformation gradient and Jacobian of the transformation. It is obvious from Eq. (5) that the transformed elastic tensor **C** is no longer to possess minor symmetry as **F** does not equally act on the subscripts, and the stress **σ** is no longer symmetric as well. However, the major symmetry still holds.

Without loss of generality, we will in the following focus on a typical case of transformation which is rotation free, i.e. a volume element undergoes only stretch, the deformation gradient is symmetric and expressed for 2D case as

$$\mathbf{F} = \delta_1 \mathbf{e}_1 \mathbf{e}_1 + \delta_2 \mathbf{e}_2 \mathbf{e}_2 \tag{6}$$

where $\delta_1$ and $\delta_2$ are the stretching ratios along the two orthogonal directions $\mathbf{e}_1$ and $\mathbf{e}_2$, respectively. The transformations corresponding to simple stretching and cylindrical cloak fall into this category. For the AMMs adhered to these transformations, their peculiar wave property and design criteria are examined in the following, which are useful for the next section of microstructure design. Assuming the virtual domain is isotropic and defined by $C_{ijkl}^{sym} = \lambda_0 \delta_{ij} \delta_{kl} + \mu_0 (\delta_{jk} \delta_{il} + \delta_{ik} \delta_{jl})$, from Eq. (5) the constitutive relation of the AMM is expressed in Voigt form as

$$\begin{pmatrix} \sigma_{11} \\ \sigma_{22} \\ \sigma_{12} \\ \sigma_{21} \end{pmatrix} = \begin{bmatrix} C_{1111} & C_{1122} & 0 & 0 \\ C_{1122} & C_{2222} & 0 & 0 \\ 0 & 0 & C_{1212} & C_{1221} \\ 0 & 0 & C_{1221} & C_{2121} \end{bmatrix} \begin{pmatrix} \varepsilon_{11} \\ \varepsilon_{22} \\ \varepsilon_{12} \\ \varepsilon_{21} \end{pmatrix} = \begin{bmatrix} \frac{\delta_1}{\delta_2}(\lambda_0 + 2\mu_0) & \lambda_0 & 0 & 0 \\ \lambda_0 & \frac{\delta_2}{\delta_1}(\lambda_0 + 2\mu_0) & 0 & 0 \\ 0 & 0 & \frac{\delta_1}{\delta_2}\mu_0 & \mu_0 \\ 0 & 0 & \mu_0 & \frac{\delta_2}{\delta_1}\mu_0 \end{bmatrix} \begin{pmatrix} u_{1,1} \\ u_{2,2} \\ u_{2,1} \\ u_{1,2} \end{pmatrix} \tag{7}$$

and the density is given by

$$\rho = \frac{1}{\delta_1 \delta_2} \rho_0. \tag{8}$$

It is seen that the transformed AMM is orthotropic with anisotropy of the in-plane shear moduli ($C_{1212} \neq C_{2121}$) along the two principal directions, impossible for a traditional orthotropic material since there is only one in-plane shear modulus. Assuming an harmonic wave with time dependence $e^{-i\omega t}$, Eqs.(4)$_2$ and (7) give the wave equation of the considered AMM

$$\begin{cases} -\omega^2 \rho u = C_{1111} u_{,xx} + C_{2121} u_{,yy} + (C_{1122} + C_{1221}) v_{,xy} \\ -\omega^2 \rho v = C_{1212} v_{,xx} + C_{2222} v_{,yy} + (C_{1122} + C_{1221}) u_{,xy} \end{cases}, \tag{9}$$

where we have used $u \equiv u_1$, $v \equiv u_2$ for convenience. Admitting a plane wave solution of $\{\tilde{u}, \tilde{v}\}^T \exp[i(k_x x + k_y y)]$, the secular equation of the two wave modes reads

$$\left( \delta_1^2 k_x^2 + \delta_2^2 k_y^2 - \frac{\omega^2}{v_S^2} \right) \left( \delta_1^2 k_x^2 + \delta_2^2 k_y^2 - \frac{\omega^2}{v_P^2} \right) = 0, \tag{10}$$

where $v_S = (\mu_0/\rho_0)^{1/2}$ and $v_P = (\lambda_0 + 2\mu_0/\rho_0)^{1/2}$ are the shear and longitudinal wave speeds for the isotropic virtual medium, respectively. The polarizations of the two wave modes in accordance to Eq. (10) are $\tilde{v}/\tilde{u} = (\delta_2 k_y)/(\delta_1 k_x)$ for P-dominated mode and $\tilde{v}/\tilde{u} = -(\delta_1 k_x)/(\delta_2 k_y)$ for S-dominated mode, respectively. The waves are pure P or S along the $k_x$ or $k_y$ direction, and both S and P waves have different phase speeds along the two directions. Graphically, the iso-frequency contour (IFC) of the orthotropic AMM features unique nested ellipses with the same aspect ratio, distinct from that of a traditional orthotropic material.

Fig. 2(a) shows a schematic scenario of wave refraction from the lower isotropic medium to the upper AMM obtained by a stretching transformation, i.e. ($y = 2y'$, $x = x'$) and ($\delta_1 = 1$, $\delta_2 = 2$). For the upper domain, the IFCs of P- and S-dominated waves are ellipses with an aspect ratio 2, while for the lower isotropic domain, the IFCs of P- and S-waves are circles. The figure shows an incident S-wave ($45°$) from the bottom-left with a polarization denoted by short red arrows. When transmitted to the AMM the wave vector is refracted to $63.4°$, while the polarization remains identical in both domains. This implies that for an incident S-wave, only a transmitted S-dominated mode is needed to satisfy the displacement and stress continuity at the interface, and similarly for P wave incidence. In other words, P and S waves are able to non-reflectively and independently be accepted and transmitted by a transformation induced AMM. It is this unique wave property to make the AMM perfectly matches elastic waves, which is extremely useful for elastic wave control. Note that for a wave beam representing the energy flux, the path will trace the group velocity denoted by $\mathbf{c}_g$. This example of Fig. 2(a) will be verified in Section 3.3 by the proposed microstructural model. For comparison, Fig. 2(b) shows the IFC of a traditional 2D orthotropic material. It is seen that only the P-mode shows an elliptical-like IFC, while the S-mode propagates at the same speeds along the two principal directions since there is only one shear modulus for a traditional 2D material.

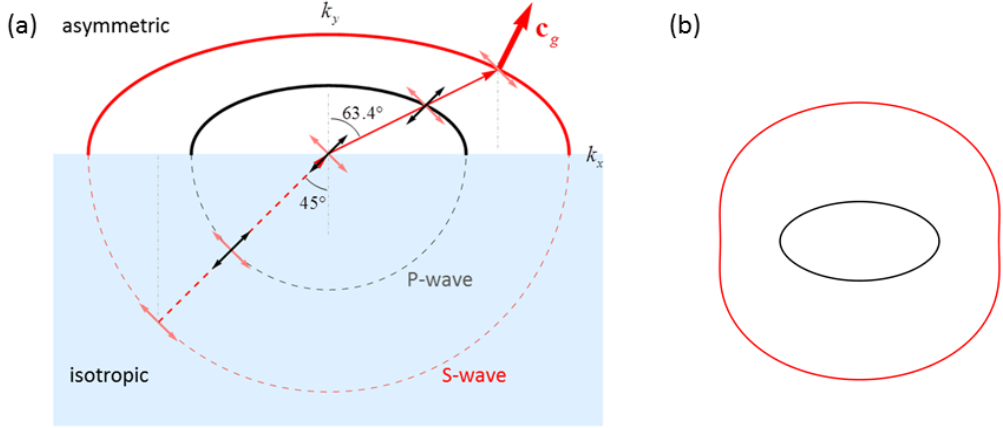

Fig.2. (a) Schematic diagram of wave refraction between an isotropic symmetric material and an AMM induced by stretching transformation; (b) IFC of a traditional orthotropic material.

## 3. A resonance based AMM model

*3.1 The lattice model and homogenization*

We propose in this section a lumped parameter model which can effectively meet the asymmetric and anisotropic properties ascribed by Eqs. (7) and (8). Fig. 3 shows the proposed unit cell of the discrete mass-spring model of AMM. The material is a periodic rectangular lattice with lattice vectors ($\mathbf{a}_1$, $\mathbf{a}_2$) thus its effective property is orthotropic. Mass points $m$ with their displacements are the observable host, while a finite sized rigid body with mass $M$ and rotation inertia $J$ is introduced within each unit cell acting as the hidden inclusion, i.e. its displacement and rotation are unobservable. The mass points are linked by horizontal and vertical springs with the same stiffness $K$, and each central rigid bar connects to the surrounding masses through four inclined springs with constant $h$. The angle $\theta$ can be generally arbitrary, however, here for simplicity we fix $\theta = \pi/4$ so that the resultant restoring force magnitude as well as the translational resonance state of the bar inclusion will be the same when displaced in any direction, hence the effective density is isotropic as required by Eq. (5). Further, as long as the lattice constant is not equal, i.e. $a_1 \neq a_2$, there will be a restoring torque between the bar and the surrounding host, hence the rotational resonance could be produced and in turn cause the observable constitutive behavior asymmetric.

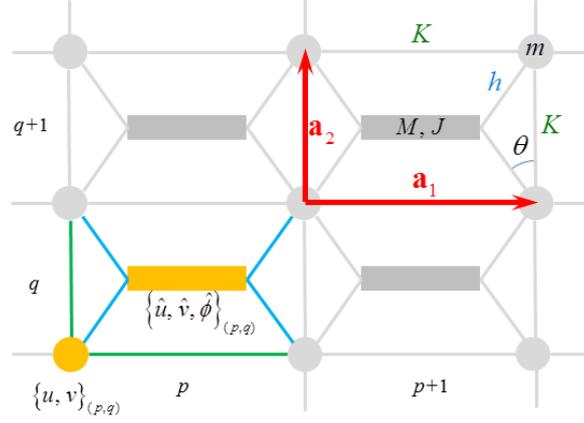

Fig. 3. The configuration of the proposed periodical lattice producing orthotropic and asymmetric elastic behavior. The lattice composes of point masses, springs and finite sized rigid body. A unit cell is highlighted with colors, cells and their DOFs are indexed by the row and column number ($p$, $q$).

The dispersion property of this structure is quite complex since both translational and rotational resonances are present. To get a more reliable estimation of the effective property of the lattice, we adopt here the homogenization procedure based on the field variable expansion (Liu et al. 2012). As shown in Fig.3, Let $\mathbf{u}_{p,q} = \{u_{p,q}, v_{p,q}\}^{\mathrm{T}}$ and $\hat{\mathbf{u}}_{p,q} = \{\hat{u}_{p,q}, \hat{v}_{p,q}, \hat{\phi}_{p,q}\}^{\mathrm{T}}$ denote the motion of the observable point mass and hidden rigid bar belonging to the unit cell ($p$, $q$), respectively, where $\hat{u}$ and $\hat{v}$ are the displacements of the bar centroid and $\hat{\phi}$ is its rotation. For the configuration (here $a_1 > a_2$), the infinitesimal displacement vectors of the left and right ends of the bar are determined as

$$\mathbf{u}^L_{p,q} = \{\hat{u}, \hat{v} - |a_1 - a_2|\hat{\phi}/2\}^{\mathrm{T}}_{p,q}, \quad \mathbf{u}^R_{p,q} = \{\hat{u}, \hat{v} + |a_1 - a_2|\hat{\phi}/2\}^{\mathrm{T}}_{p,q}. \tag{11}$$

For cell ($p$, $q$), the potential energy relevant to the two host springs and four internal springs can be written as

$$U_{p,q} = \frac{1}{2}K\left[\left(u_{p+1,q} - u_{p,q}\right)^2 + \left(v_{p,q+1} - v_{p,q}\right)^2\right] + \frac{1}{2}h\left[\left(\mathbf{e}' \cdot \mathbf{u}^L_{p,q} - \mathbf{e}' \cdot \mathbf{u}_{p,q}\right)^2 + \left(\mathbf{e}'' \cdot \mathbf{u}^L_{p,q} - \mathbf{e}'' \cdot \mathbf{u}_{p,q+1}\right)^2 + \left(\mathbf{e}'' \cdot \mathbf{u}^R_{p,q} - \mathbf{e}'' \cdot \mathbf{u}_{p+1,q}\right)^2 + \left(\mathbf{e}' \cdot \mathbf{u}^R_{p,q} - \mathbf{e}' \cdot \mathbf{u}_{p+1,q+1}\right)^2\right], \tag{12}$$

where $\mathbf{e}' = (1/\sqrt{2}, 1/\sqrt{2})^{\mathrm{T}}$ and $\mathbf{e}'' = (1/\sqrt{2}, -1/\sqrt{2})^{\mathrm{T}}$ are unit vectors representing the directions of the internal spring, and the kinetic energy is given by

$$K_{p,q} = \frac{1}{2}m|\dot{\mathbf{u}}_{p,q}|^2 + \frac{1}{2}M\left(\dot{\hat{u}}^2_{p,q} + \dot{\hat{v}}^2_{p,q}\right) + \frac{1}{2}J\dot{\hat{\phi}}^2_{p,q}. \tag{13}$$

The Hamiltonian of the system is therefore

$$H = \sum_{p,q}\left[U_{p,q} + K_{p,q}\right]. \tag{14}$$

By using Hamilton's principle, the harmonic wave equations of the infinite periodic lattice are obtained for the observable $\mathbf{u}_{p,q}$ as,

$$\omega^2 m u_{p,q} = \frac{\partial H}{\partial u_{p,q}}, \qquad \omega^2 m v_{p,q} = \frac{\partial H}{\partial v_{p,q}}, \tag{15}$$

and for the hidden $\hat{\mathbf{u}}_{p,q}$ as

$$\omega^2 M \hat{u}_{p,q} = \frac{\partial H}{\partial \hat{u}_{p,q}}, \qquad \omega^2 M \hat{v}_{p,q} = \frac{\partial H}{\partial \hat{v}_{p,q}}, \qquad \omega^2 J \hat{\phi}_{p,q} = \frac{\partial H}{\partial \hat{\phi}_{p,q}}. \tag{16}$$

The right hand sides (RHS) of Eqs. (15) and (16) are lengthy and not reported here. Generally, for the observable DOF, RHS of Eq. (15) are relative to both the observable and hidden variables of the cell ($p$, $q$) and its surroundings, while RHS of Eq. (16) are only relative to the observable variables of the surrounding cells since the bar inclusions are isolated in each cell and there is no direct interaction of two bar inclusions. Before replacing the discrete system with a homogenized media, we eliminate the hidden DOFs and retain Eq. (15) and the host variables only. This can be done by solving Eq. (16) for $\hat{\mathbf{u}}_{p,q}$, and substituting it into Eq. (15). The process is tedious but straightforward, and finally the discrete governing equations represented by only the observable host DOFs are obtained as

$$\begin{aligned}(2h - m\omega^2) u_{0,0} &= K(u_{-1,0} - 2u_{0,0} + u_{+1,0}) + \frac{hT(\omega)}{4}(u_{-1,-1} + u_{-1,+1} + 4u_{0,0} + u_{+1,-1} + u_{+1,+1} \\ &\quad + v_{-1,-1} - v_{-1,+1} - v_{+1,-1} + v_{+1,+1}) - \frac{hR(\omega)}{8}(u_{-1,-1} - 2u_{-1,0} + u_{-1,+1} + 2u_{0,-1} - 4u_{0,0} + 2u_{0,+1} \\ &\quad + u_{+1,-1} - 2u_{+1,0} + u_{+1,+1} + v_{-1,-1} - v_{-1,+1} - v_{+1,-1} + v_{+1,+1})\end{aligned} \tag{17}$$

$$\begin{aligned}(2h - m\omega^2) v_{0,0} &= K(v_{0,-1} - 2v_{0,0} + v_{0,+1}) + \frac{hT(\omega)}{4}(u_{-1,-1} - u_{-1,+1} - u_{+1,-1} + u_{+1,+1} + v_{-1,-1} \\ &\quad + v_{-1,+1} + 4v_{0,0} + v_{+1,-1} + v_{+1,+1}) - \frac{hR(\omega)}{8}(u_{-1,-1} - u_{-1,+1} - u_{+1,-1} + u_{+1,+1} + v_{-1,-1} + 2v_{-1,0} \\ &\quad + v_{-1,+1} - 2v_{0,-1} - 4v_{0,0} - 2v_{0,+1} + v_{+1,-1} + 2v_{+1,0} + v_{+1,+1})\end{aligned} \tag{18}$$

where we have set $p = q = 0$ as the referential cell, and two functions of the frequency $\omega$

$$T(\omega) = \frac{\omega_T^2}{\omega_T^2 - \omega^2}, \quad R(\omega) = \frac{\omega_R^2}{\omega_R^2 - \omega^2}, \tag{19}$$

are introduced with $\omega_T = \sqrt{2h/M}$ and $\omega_R = \sqrt{h(a_1 - a_2)^2/(2J)}$ being the frequencies related to the translational and rotational resonances of the bar inclusion, respectively. As long as the background wave length is large enough, it is reasonable to represent the adjacent $\mathbf{u}_{\pm 1,\pm 1}$ using Taylor's expansion with respect to $\mathbf{u}_{0,0}$, specifically, up to the second order, we have for the displacement $u$

$$u_{0,0} \equiv u, \qquad dx_{\pm 1} = \pm a_1, \qquad dy_{\pm 1} = \pm a_2,$$

$$u_{\pm 1,\pm 1} = u + u_{,x}dx_{\pm 1} + u_{,y}dy_{\pm 1} + \frac{1}{2}u_{,xx}dx_{\pm 1}^2 + \frac{1}{2}u_{,yy}dy_{\pm 1}^2 + u_{,xy}dx_{\pm 1}dy_{\pm 1}, \tag{20}$$

and the similar expansion for $v$. Substituting Eq. (20) into Eqs. (17) and (18) turns out that the obtained continuous equations match exactly the form of an orthotropic AMM, i.e. Eq. (9). By comparing the coefficients we read off directly the effective density and the four effective stiffness constants as

$$\rho = \frac{1}{a_1 a_2}\left[m + MT(\omega)\right] \tag{21}$$

$$C_{1111} = \frac{a_1}{a_2}\left[K + \frac{h}{2}T(\omega)\right], \qquad C_{2222} = \frac{a_2}{a_1}\left[K + \frac{h}{2}T(\omega)\right]$$

$$C_{1212} = \frac{h}{2}\frac{a_1}{a_2}\left[T(\omega) - R(\omega)\right], \qquad C_{2121} = \frac{h}{2}\frac{a_2}{a_1}\left[T(\omega) - R(\omega)\right] \tag{22}$$

and a summation of two constants

$$C_{1122} + C_{1221} = \frac{h}{2}\left[2T(\omega) - R(\omega)\right]. \tag{23}$$

To split this term, we notice that the lattice is not statically stable hence the shear modulus $C_{1221}$ has to vanish when $\omega \to 0$. Meanwhile, it can be anticipated that if the aspect ratio of the unit cell tends to unity, i.e. $a_1 = a_2$, the material will effectively reduce to a symmetric material, i.e. $C_{1212} = C_{2121} = C_{1221}$. With these arguments in mind and taking Eq. (22) into account, it is natural to have the two effective constants as

$$C_{1122} = \frac{h}{2}T(\omega), \qquad C_{1221} = \frac{h}{2}\left[T(\omega) - R(\omega)\right]. \tag{24}$$

It is seen from Eqs. (21) - (24) that all the effective constants are frequency dependent. In particular, the density and normal moduli $C_{1111}$, $C_{2222}$, $C_{1122}$ are relative to the translational resonance ($T(\omega)$) only, while the shear moduli are relative to both the translational and rotational resonances.

It is observed that the medium are effectively asymmetric ($C_{1212} \neq C_{1221}$) providing that $a_1 \neq a_2$. Referring to Eq. (7), we see that the effective stiffness parameters of the proposed lattice satisfy the relations of those of a pure stretching transformation induced

AMM, say $C_{1111} C_{2222} = (C_{1122} + 2 C_{1221})^2$ and $C_{1212} C_{2121} = C_{1221}^2$. It is also very interesting to see that, to conform to the stretching transformation, the ratio of the stretching on the two principal directions is just the aspect ratio of the unit cell.

$$\frac{a_1}{a_2} = \frac{\delta_1}{\delta_2}. \tag{25}$$

*3.2 Parametrization with respect to stretching transformation*

The proposed model effectively manifests shear stress asymmetry near the resonance, the next question is whether it can realize the wave functionality defined by a transformation method. Concretely, given a background isotropic elastic medium ($\lambda_0$, $\mu_0$, $\rho_0$), a transformation defined by stretching ratios ($\delta_1$, $\delta_2$) along the two principal directions as well as a desired operation frequency $\omega_{ext}$, we should determine the microstructural parameters ($K$, $h$, $m$, $M$, $J$) so that the effective properties at $\omega=\omega_{ext}$ meet the those of transformation media, i.e. Eq. (7). Because of Eq. (25), the rectangular unit cell can be considered as stretched in the same scale from a square cell (with size *a*) in the virtual domain, i.e. $a_1 = \delta_1 a$, $a_2 = \delta_2 a$. Define $N$ the number of unit cells within a single wave length $\lambda_S$ of the S-wave of the background medium

$$\frac{a_1 a_2}{\delta_1 \delta_2} N^2 = a^2 N^2 = \lambda_S^2 = \frac{4\pi^2}{\omega_{ext}^2} \frac{\mu_0}{\rho_0}. \tag{26}$$

Since the unit cell and wave length are stretched synchronously in the transformed space, sufficiently large $N$ (usually $N > 10$) ensures the long wave assumption of the homogenization. For convenience, we further introduce a parameter $\xi$ as the ratio between frequency squares of the rotation and translational resonances

$$\xi = \omega_R^2 / \omega_T^2 = \frac{a^2 M}{4J} (\delta_1 - \delta_2)^2, \tag{27}$$

physically it indicates the two kinds of inertia of the bar inclusion. Comparing Eqs. (21) - (24) with Eq. (7), the following for relations are obtained

$$\frac{4\pi^2}{\omega_{ext}^2 N^2} \mu_0 = m + MT(\omega_{ext}), \qquad \lambda_0 = \frac{h}{2} T(\omega_{ext}),$$

$$\lambda_0 + 2\mu_0 = K + \frac{h}{2} T(\omega_{ext}), \qquad \mu_0 = \frac{h}{2} T(\omega_{ext}) - \frac{h}{2} R(\omega_{ext}). \tag{28}$$

with which the microscopic masses and springs can be solved as

$$K = 2\mu_0, \qquad h = \frac{2(\xi-1)\lambda_0(\lambda_0-\mu_0)}{(\xi-1)\lambda_0+\mu_0},$$

$$M = \frac{4\xi(\xi-1)\lambda_0\mu_0(\lambda_0-\mu_0)}{[(\xi-1)\lambda_0+\mu_0]^2 \omega_{ext}^2}, \qquad m = \frac{4\mu_0}{\omega_{ext}^2}\left[\frac{\pi^2}{N^2} - \frac{\xi\lambda_0}{(\xi-1)\lambda_0+\mu_0}\right]. \tag{29}$$

It is remarked that although the microscopic mass and spring constants can be obtained from Eq. (29) anyway, physically theses constants must be positive, which in fact places certain constraints on the background material. To see this, it is more convenient to express the above equations using Young's modulus $E_0$ and Poisson's ratio $v_0$ as

$$K = \frac{E_0}{1+v_0}, \qquad h = \frac{2E_0}{(1+v_0)(1-2v_0)} \frac{(\xi-1)v_0(4v_0-1)}{(1+2\xi v_0-4v_0)},$$

$$M = \frac{4\xi E_0}{\omega_{ext}^2(1+v_0)} \frac{(\xi-1)v_0(4v_0-1)}{(1+2\xi v_0-4v_0)^2}, \qquad m = \frac{2E_0}{\omega_{ext}^2(1+v_0)}\left[\frac{\pi^2}{N^2} - \frac{2\xi v_0}{1+2\xi v_0-4v_0}\right]. \tag{30}$$

Requiring the above expressions all positive, the following three conditions can be obtained,

$$\begin{aligned}
&cond.1: \quad (\xi-1)v_0(4v_0-1) > 0, \\
&cond.2: \quad 1 + 2\xi v_0 - 4v_0 > 0, \\
&cond.3: \quad \frac{\pi^2}{N^2} - \frac{2\xi v_0}{1+2\xi v_0 - 4v_0} > 0,
\end{aligned} \tag{31}$$

which limit the possible background Poisson's ratio with specific choices of the bar inclusion inertia feature $\xi$ and the cell division $N$ per wavelength.

Fig. 4 plots individually the allowable regions of the above three conditions and that of the combination of the three conditions, respectively, with $N = 10$. It is noticed that when $\xi > 1$, i.e. the rotational resonance frequency is higher than the translational one, only the background media with negative Poisson's ratio can be realized with the model, conversely the background media with positive Poisson's ratio requires that $\xi < 1$. Moreover, mainly due to the constraint of the *condition* 3, where $N = 10$ is chosen, for positive $v_0$ the allowable region of parameter combination is severely limited, and not full range of $v_0$ can be covered (about $0 < v_0 < 0.25$), while for negative $v_0$ wider parameter choice can be made and full range of $v_0$ can be realized ($-1 < v_0 < 0$). Eq. (30) or (29) together with the diagram in Fig.4 set up an inverse design scheme that output the microstructural parameters for a given background medium and stretching transformation.

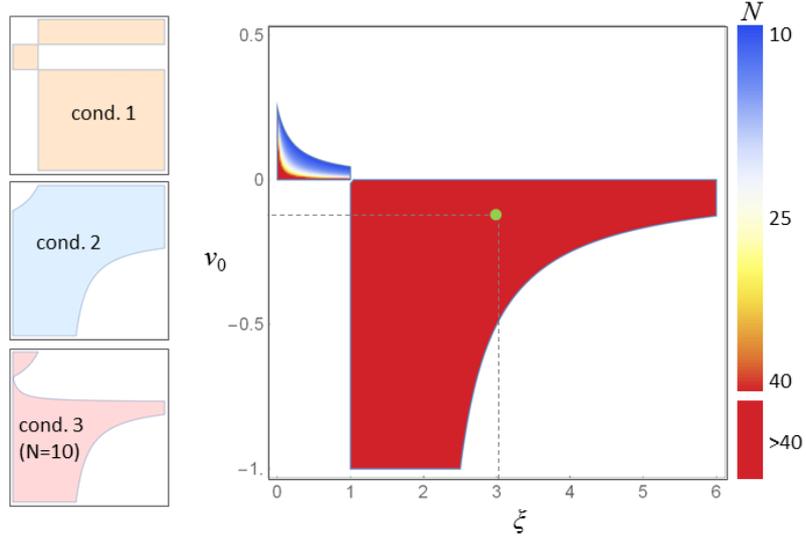

Fig. 4. The allowable combination of $v_0$ and $\xi$ restricted individually by *condition* 1, 2 and 3 with $N=10$, respectively (left panel), and that simultaneously restricted by the three condition (right panel). The color contour in the right panel indicates the maximum $N$ that can be used.

Next, let us take an example to examine to what extent the homogenized AMM can characterize the wave behavior of the lattice. Here $v_0 = -0.1$, $\xi = 3$ are adopted as marked by the green dot in Fig. 4, and the cell division $N = 10.47$ is chosen. Other parameters are $E_0 = 5$, $\rho_0 = 1$, $\delta_1 = 1$, $\delta_2 = 2$, and the operating frequency is $\omega_{\text{ext}} = 1$. Or equivalently, the background isotropic material is defined by

$$\rho_0 = 1, \quad \lambda_0 = -0.4630, \quad \mu_0 = 2.7778, \tag{32}$$

and the microstructural parameters are evaluated using Eq. (30) as

$$K = 5.556, \quad h = 3.241, \quad M = 29.167, \quad m = 9.333, \quad a_2 = 2a_1 = 2, \tag{33}$$

with which the two resonance frequencies are evaluated as $\omega_T = 0.471$ and $\omega_R = \xi^{1/2}\omega_T = 0.817$, respectively. It is verified through Eqs. (21)-(24) that the effective parameters exactly agree with Eq. (7) at the operating frequency $\omega_{\text{ext}} = 1$.

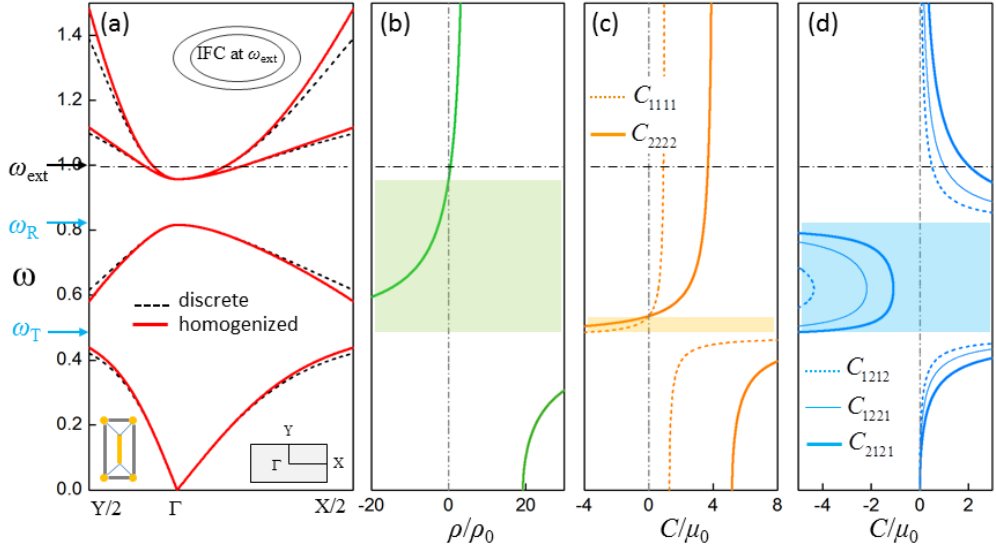

Fig. 5. (a) Band diagram calculated from the discrete and the homogenized model, where the insets are: the unit cell (bottom-left), the first Brillouin zone (bottom right) and the IFC at the operation frequency (upper); (b) effective density, (c) effective longitudinal moduli and (d) effective shear moduli as the function of frequency, the colored zones indicate where the effective parameters are negative.

In Fig. 5(a) the band diagram calculated using Eqs. (9) and (21)-(23) are graphed by solid curves, while the accurate dispersion obtained by the Bloch wave solution of the discrete wave Eqs. (17) and (18) are also accompanied by dotted curves for comparison. Fig. 5(b)-(d) show the effective density, longitudinal moduli and shear moduli, respectively, as the function of frequency $\omega$. It is seen that the dispersion curves calculated from the homogenized medium match quite well the accurate solution for small wave numbers for all the branches and the discrepancy goes large as the 2$^{nd}$ order Taylor's expansion is not enough when the background wave length in the host is short. Fig. 5(d) shows that near $\omega_T$ and $\omega_R$, the effective shear moduli demonstrate a significant asymmetry ($C_{1212} \neq C_{1221}$) and anisotropy ($C_{1212} \neq C_{2121}$) while they go to zero away from the resonance. There are totally five branches in accordance to the five DOFs within a unit cell. There are actually two acoustic bands in the frequency region $\omega \in [0, \omega_T]$, since the lattice is not statically stable, the first (acoustic shear) branch overlaps with the horizontal axis and is not shown here. For the region $\omega \in [\omega_T, \omega_R]$, only one branch with negative slope is found, indicating the opposed sign of the group and phase speeds. Referring to Fig. 5(b) and (d), in this frequency region both the effective density and shear moduli are negative while the longitudinal moduli are positive, hence this negative branch is shear dominated. Finally, when $\omega > \omega_R$, both the P- and S- modes are present since all the positive properties can be found and the asymmetric elastic wave behavior is very pronounced. Therefore in this region the proposed lattice is possible to meet the functions of transformation of elastic wave, providing that the wave vector is sufficiently near to the Brillouin zone center. In particular, the IFC extracted directly from the Bloch wave analysis at the targeted operation frequency $\omega_{ext} = 1$ is displayed in the upper inset of Fig. 5(a), in which obvious nested ellipses with the same aspect ratio 2 are observed, in accordance with the given stretching ratio.

*3.3 PML obtained by a simple stretching transformation*

For the example of the microstructure design used in Fig.5, the stretching ratios and the IFC shown in the inset are the same with that in Fig. 2(a) with continuum model, therefore the wave behavior of the lattice is expected to follow the same prediction. In this subsection, we will examine whether the lattice model of AMM can achieve an elastic PML defined by a simple stretching transformation through a full-wave finite element (FE) simulation. As shown in Fig. 6(a), the FE model consists of an elastic continuum domain (Eq. (32)) sandwiched by a slab made of 50× 6=300 discrete unit cells with the microstructural parameters given by Eq.(33). The interface connectivity is ensured by proper meshing and sharing nodes of the solid elements with the discrete masses, and absorbing boundary are used in the surroundings of the simulated domain to eliminate the reflecting wave. An S-wave Gaussian beam with $\omega_{ext}=1$ and at an incident angle 45° is launched from the bottom-left. The numerical simulations in this paper are implemented and solved with hybrid FE models consisting of solid, mass, spring, rigid body and absorbing layer elements developed in Mathematica software package.

Fig. 6(a) shows the field of the displacement magnitude. It is seen that the wave beam is perfectly transmitted through the micro-structured slab without any reflection, meanwhile the outgoing beam offsets as if the wave path is also stretched to follow the domain, the marked energy flow (green dotted) in the discrete region agrees very good with the group velocity in Fig. 2(a). For the wave vector in the AMM, it can be identified from the zoomed plot shown by Fig. 6(b), where the wave front is designated by dash-dotted lines linking the masses with the same phase. The refraction angle (63.4°) obtained by simulation matches exactly that of Fig. 2(a). More importantly, it is clearly seen from Fig. 6(b) that the particle polarization (black dashed arrows) is identically along the same direction ($45°$) both in the continuum region and the discrete region. Similar prefect transmission also happens for a P-wave beam or even point source excitation. This example proves that our concept to make the AMM is valid and efficient for elastic wave control. In the next section, we will employ it to design an elastic wave cloak.

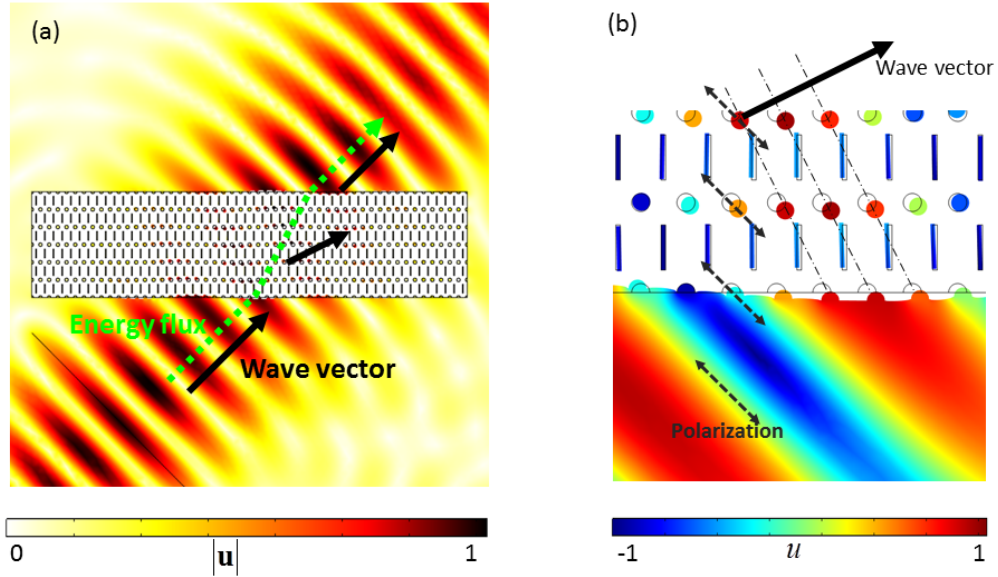

Fig. 6 (a) Perfect transmission of an S-wave Gaussian beam incident on a latticed AMM slab corresponding to uniform stretching along the vertical direction, colors indicate displacement magnitude $|\mathbf{u}|=(u^2+v^2)^{1/2}$; (b) zoomed plot around the refraction interface, dash-dotted lines mark the wave front, gray circles and dashed arrows denote the original positions and the polarization, respectively, colors indicate the displacement component $u$. Springs are not shown here.

## 4. Application to design an elastic cylindrical cloak

Now consider a 2D cylindrical cloak as shown in Fig. 7(a), under an axisymmetric transformation $r' = f(r) = b_1(r-b_0)/(b_1-b_0)$, $\theta' = \theta$, the virtual domain $0 < r' < b_1$ is squeezed into the cloak $b_0 < r < b_1$, and in the polar coordinate system $\mathbf{F} = \delta_r\,\mathbf{e}_r\mathbf{e}_r + \delta_\theta\,\mathbf{e}_\theta\mathbf{e}_\theta$ where the $r$-dependent stretching ratios along $r$ and $\theta$ direction are respectively

$$\delta_r(r) = \left(\frac{\mathrm{d}f}{\mathrm{d}r}\right)^{-1} = \frac{b_1-b_0}{b_1}, \qquad \delta_\theta(r) = \frac{r}{f} = \frac{(b_1-b_0)r}{b_1(r-b_0)}. \tag{34}$$

Since $\delta_\theta/\delta_r>1$, the latticed cloak is of a gradient pattern that the bar inclusions are along the $\theta$-direction, as shown in Fig. 7(b). Considering the cloak is divided into $N_{sec}$ sectors of unit cells in the circumferential direction, and the cell layers are indexed inwards and starting from 1 for the outermost layer. For the layer $n$, the aspect ratio of a unit cell is evaluated at its outer radius $r_n$ using Eq. (34), i.e., $\delta_r/\delta_\theta = (r_n - b_0)/r_n$. Supposing $r_n$ is known, an edge size of the unit cell is $a_\theta(r_n) = 2\pi r_n / N_{sec}$, hence the other edge size, i.e. the thickness of layer $n$ is determined as $a_r(r_n) = a_\theta \delta_r/\delta_\theta = 2\pi(r_n - b_0)/N_{sec}$. For the next layer $r_{n+1} = r_n - a_r(r_n)$, and notice that $r_1 = b_1$ for layer 1, the unit cell sizes for each layer of the latticed cloak can be recursively figured out. Then the masses and springs constants of the unit cells can be obtained using Eqs. (26) and (29), from which it is noticed that only the mass $m$ differs for different layers and other parameters are the same for the whole lattice. Since the rectangular unit cells are embedded into the cylindrical frame, enough sector division $N_{sec}$ should be chosen so that the unit cells do not distort too much. The sector division should also guarantee the long wave assumption. Since for the outermost layer $\delta_\theta=1$, using Eq. (26) a relation can be obtained as $Nb_1\omega_{ext}=N_{sec}(\mu_0/\rho_0)^{1/2}$, which can be used as an additional check. Finally, because that the unit cell aspect ratio $\delta_\theta/\delta_r$ tends to infinity approaching the inner boundary of the cloak, theoretically infinite number of layers are needed for the latticed cloak to reach strictly the inner radius $b_0$. However, in practice the number of layer can be truncated to an acceptable value and the cloaking performance is still good.

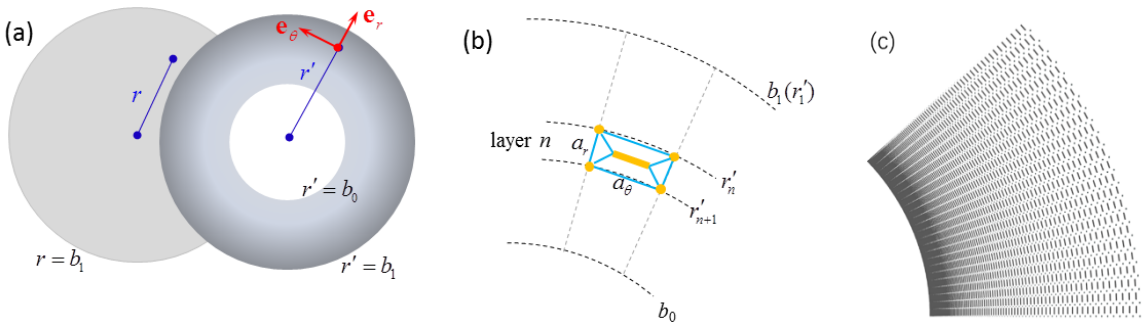

Fig. 7 (a) Transformation of a cylindrical cloak; (b) assembling unit cells to the latticed cloak; (c) a portion of the latticed cloak.

To verify the previous design procedure and the applicability of the proposed AMM lattice model, we built an illustrative latticed cloak and performed full wave FE simulation, as shown in Fig. 8. Here the host isotropic media are characterized by Eq. (32). The cloak region is bounded by outer radius $b_1=1$ and inner radius $b_0=0.5$, and is divided into $N_{sec}=200$ sectors. The layer number is truncated to 100 so that the assembly totally contains 20000 unit cells. A portion of the final latticed cloak is shown in Fig. 7(c). The bar inclusion inertia feature $\xi = 3$ is chosen for all cells, and the operating frequency is $\omega_{ext} = 2\pi$. At the outer boundary, the latticed cloak is coupled with elastic continuum background in the same manner with the previous example, while the inner boundary is set as free thus the cloaked region is a void. To estimate the cloaking performance, we consider the same cloak is illuminated by incident plane harmonic P and S-wave separately. Incident plane waves with Gaussian profiles are launched from the left boundary of the background domain, and absorbing boundaries are attached to all sides of the simulation domain to eliminate the wave reflection.

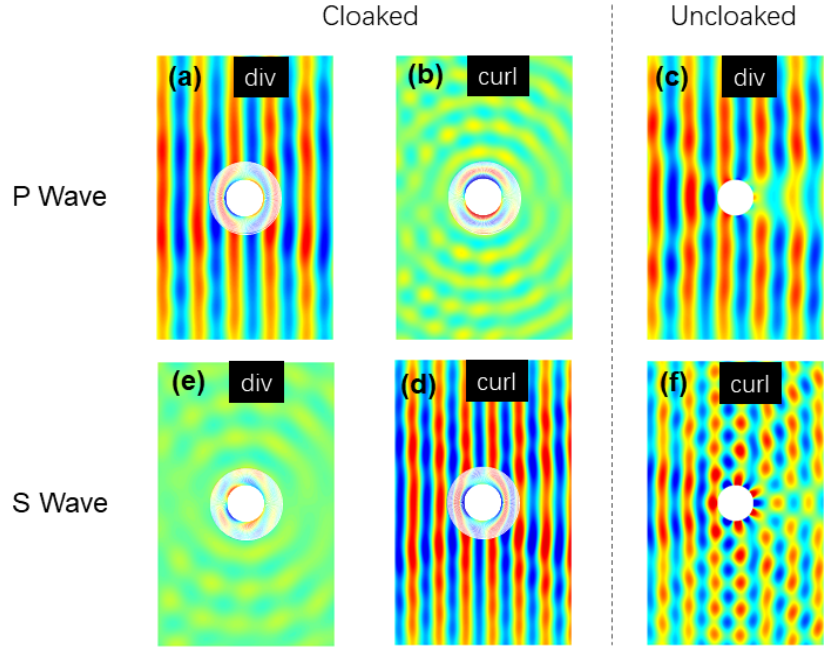

Fig. 8 Displacement (a) divergence and (b) curl plots for cloaked void and (c) divergence plot for uncloaked void in case of P wave incidence; displacement (e) divergence and (d) curl plots for cloaked void and (f) curl plot for uncloaked void in case of S wave incidence.

The simulation results are shown in Fig. 8, where the first and second rows plot the scattering fields for the incident P and S waves, respectively. For reference, the third column of the figure shows the results of uncoated cavity under the same wave illumination. The wave scatterings are shown as color contours indicating the real part of the divergence or curl of the displacement field, which is to check out the P or S wave contents. Note that in the discrete cloak shell only the mass displacements are interpolated and shown. For the P-wave incidence, the divergence and curl fields are shown by Fig. 8(a) and (b), respectively. Since the cloaked void is intend to mimic intact domain without scattering, the expected ideal results would separately be undisturbed plane wave for divergence and zero field for curl. Due to the imperfectness of the latticed cloak, slight wave front distortion is observed in divergence plot, also small amount of shear wave due to scattering is found by the curl plot. However, comparing Fig. 8(a) with the severely disturbed divergence plot for the uncloaked case (Fig. 8(c)), the cloaking effect is very obviously observed. For the S-wave incidence with shorter wavelength, as depicted by Fig. 8(d)-(f), similarly good cloaking effect is also observed except that the dominated field is the curl plot. The imperfectness found for the lattice cloak relative to the ideal one can be attributed to several sources, e.g. the error between the homogenized and real lattice, the truncation of the number of layers so that the cloak cannot reach its theoretical inner radius, the non-smooth interaction between the discrete masses and the continuum domain, and finally the unavoidable distortion of rectangular unit cell to a trapezoidal shape in building up the cylindrical cloak.

## 5. Conclusion

In this study, we proposed a resonance-based 2D metamaterial model for which the minor symmetry of the elastic tensor can be broken, without introducing external body torque or any other active methods. It turns out that the asymmetry and anisotropy of the model are able to be tuned to meet those of the transformation induced material so that full control of elastic wave is brought to reality. Complete set of design tools including the homogenization and the inverse determination of the microstructural parameters from a given background media and transformation are analytically given, they are validated by the full wave FE simulation through designing an elastic PML and an elastic cloak, respectively. The AMM model is purely passive and the loss of minor symmetry stems from the out-of-phase rotation of the hidden inclusions relative to the observable masses. The skew-symmetric part of the overall stress, which represents the micro-moment acting on each unit cell, is self-balanced dynamically with the rotational inertia of the inclusion.

Though the proposed AMM model enlightens a practical way for full control of elastic wave, at present many problems still remain open towards its application. The proposed model is only applicable for non-rotational transformation for which the transformation gradient **F** can be diagonalized, and new types of microstructure is to be discovered for more general coordinate transformation. The achievable medium in virtual space is severely restricted with the present model especially for positive Poisson's ratio, whether it is a general barrier or can be bypassed by other configurations is to be clarified. Finally, works in the near future also include replacing the proof-of-concept discrete model with a continuum version for the experimental demonstration.

**Acknowledgements**

This work was supported by the National Natural Science Foundation of China (grant numbers 11632003, 11972083, 11972080, 11802017).